\documentclass[12pt]{article}
\usepackage{amsmath,amssymb}
\usepackage{graphics}
\usepackage[dvipdfm]{graphicx}

\begin{document}
\baselineskip=16pt
\begin{titlepage}
\begin{flushright}
{\small OU-HET 666/2010}\\
\end{flushright}
\vspace*{1.2cm}

\begin{center}

{\Large\bf 
TeV-scale seesaw from a multi-Higgs model
} 
\lineskip .75em
\vskip 1.5cm

\normalsize
{\large Naoyuki Haba}
and
{\large Masaki Hirotsu}

\vspace{1cm}

{\it Department of Physics, 
 Osaka University, Toyonaka, Osaka 560-0043, 
 Japan} \\

\vspace*{10mm}

{\bf Abstract}\\[5mm]
{\parbox{13cm}{\hspace{5mm}
%

We suggest new simple model of generating tiny neutrino masses  
 through a TeV-scale seesaw mechanism
 without requiring tiny Yukawa couplings.   
This model is a simple extension of the standard model  
 by introducing 
 extra one Higgs singlet, and 
 one Higgs doublet 
 with a tiny vacuum expectation value.  
Experimental constraints, electroweak
 precision data and no  
 large flavor changing neutral currents, 
 are satisfied 
 since the extra doublet only has a Yukawa interaction 
 with lepton doublets and right-handed neutrinos, and 
 their masses are heavy of order a TeV-scale. 
Since active light neutrinos are Majorana particles, 
 this model predicts a neutrinoless double beta decay. 

}}

\end{center}

\end{titlepage}

\section{Introduction}

The recent neutrino oscillation experiments 
 gradually reveal a 
 structure of 
 lepton sector\cite{Strumia:2006db, analyses}.  
However, from the theoretical point of view, 
 smallness of neutrino mass is still a mystery and 
 it is one of the
 most important clues to find new physics beyond the
 standard model (SM). 
Seesaw mechanism naturally realizes tiny  
 masses of active neutrinos 
 through heavy particles coupled with 
 left-handed neutrinos. 
In usual type I seesaw\cite{Type1seesaw}, 
 tiny neutrino masses of order 0.1 eV 
 implies an existence of 
 right-handed neutrinos with super-heavy 
 Majorana masses, 
 which are almost decoupled in the low-energy effective
 theory, and then 
 few observations are expected in collider experiments. 
Some people consider a possibility of
 reduction of seesaw scale to TeV, 
 where effects of 
 TeV-scale right-handed neutrinos might be 
 observed in collider 
 experiments such as LHC and ILC\cite{TeVseesaw, Haba:2009sd}. 
However, they must introduce a fine-tuning in order to realize  
 both tiny neutrino mass 
 and detection of the evidence of right-handed neutrinos from 
 a mixing with 
 the SM particles. 

How about considering a possibility that 
 smallness of the neutrino masses comparing to those of quarks and
 charged leptons is originating from an extra Higgs
 doublet with a tiny vacuum expectation value (VEV) of order  
 0.1 eV.  
It is an idea that 
 neutrino masses are much smaller than other fermions 
 since the origin of them comes from different VEV of different
 Higgs doublet, 
 and where  
 tiny neutrino Yukawa couplings are not required. 
This kind of model has been considered 
 in Dirac neutrino 
 case\cite{{Nandi}, {Davidson:2009ha}, {Logan:2010ag}}. 

In this paper, we would like to propose 
 a simple model for Majorana neutrino case, 
 which is a 
 renormalizable model with minimal
 extension of the SM which appears entirely below the TeV-scale.  
A similar setup was proposed firstly in Ref.\cite{Ma}, 
 where a global $U(1)$
 lepton number symmetry is violated explicitly. 
Tiny Majorana neutrino masses are obtained  
 through a TeV-scale type I seesaw mechanism 
 without requiring tiny Yukawa couplings. 
This model contains 
 extra one Higgs singlet, and 
 one Higgs doublet 
 with a tiny VEV.  
%
As for 
 extending a Higgs sector, 
 there are 
 constraints in general, which are consistency of 
 electroweak precision data\footnote{
It is pointed out that the second Higgs doublet heavier than SM-like Higgs 
 can potentially make the precision electroweak data 
 be consistent\cite{Barbieri:2006dq}. 
} 
 and 
 absence of large flavor changing neutral
 currents (FCNCs)\cite{2hdm}.  
In our model, 
 both two constraints are 
 satisfied 
 since the extra doublet only has a Yukawa interaction 
 with lepton doublets and right-handed neutrinos\footnote{
This is a kind of a ``leptonic Higgs'' which could explain 
 PAMERA and ATIC results\cite{Goh:2009wg}. 
}, and 
 their masses are heavy 
 enough to suppress FCNCs
 although its VEV is of order 0.1 eV. 
The extra Higgs doublet yields a neutral scalar and a
 neutral pseudo-scalar, and 
 a charged Higgs particles, 
 which can 
 provide collider signatures. 
This charged Higgs can contribute to the lepton flavor 
 violating processes. 
The extra singlet produces TeV-scale Majorana masses of right-handed 
 neutrinos, and yields a neutral scalar and a neutral pseudo-scalar 
 with a lepton number. 
Other phenomenology will be also represented such as 
 the charged Higgs decay processes. 
Notice that the decay of the charged Higgs to quarks and 
 charged leptons are strongly suppressed due to absence of 
 direct interactions among them, which 
 is one of different points from usual 
 two Higgs double models.  
Since active light neutrinos are Majorana particles, 
 this model predicts a neutrinoless double beta decay.

This paper is organized as follows. 
In section 2, we show a setup of this model, 
 and analyze its vacuum and mass spectra. 
In section 3, 
 we discuss some phenomenology's. 
Finally, we summarize our conclusions.

\section{A Model}

\subsection{Lagrangian and vacuum}

In our model, we introduce one doublets Higgs $H_{\nu}$ and
 one singlet Higgs $S$, which has a lepton number and couples
 with right-handed neutrinos,
 in addition to the SM. 
The SM Higgs doublet $H$ and new Higgs double $H_{\nu}$ are denoted as 
\begin{eqnarray}
H=
\left(
\begin{array}{c}
H^{0}\\
H^{-}
\end{array}
\right)
,\;\;
H_{\nu}=
\left(
\begin{array}{c}
H^{0}_{\nu}\\
H^{-}_{\nu}
\end{array}
\right). 
\end{eqnarray}
We introduce $Z_{3}$-symmetry, 
 whose charges (and also lepton number) are shown 
 as the following table. 
\begin{table}[h]
\centering
\begin{center}
\begin{tabular}{|l|c|c|} \hline
fields  &  $Z_{3}$-charge & lepton number \\ \hline\hline
SM Higgs doublet, $H$  &  1 & 0 \\ \hline
new Higgs doublet, $H_{\nu}$, which couples with $N$ 
 &  $\omega^{2}$ & 0 \\ \hline
new Higgs singlet, $S$, which has a lepton number  &  $\omega$  &
 $-2$ \\ \hline
Right-handed neutrinos, $N$  &  $\omega$  & 1 \\ \hline
Others  &  1  &  $1$: leptons, $0$: quarks \\ \hline
\end{tabular}
\end{center}
\end{table}

\vspace{-5mm}
\noindent
Under the discrete symmetry, 
 Yukawa interactions are given by 
\begin{eqnarray}
{\mathcal L}_{yukawa}=y^{u}\bar{Q}_{L}HU_{R}
 +y^d \bar{Q}_{L}\tilde{H}D_{R}+y^{l}\bar{L}\tilde{H}E_{R} 
 +y^{\nu}\bar{L}H_{\nu}N+\frac{1}{2}y^{N}S\bar{N^{c}}N +{\rm h.c.}\; 
\label{22}
\end{eqnarray}
where 
 $\tilde{H}=i\sigma_{2}H^{\ast}$.  
We omit a generation index here. 

Through the interactions of 
 $y^{\nu}\bar{L}H_{\nu}N + \frac{1}{2}y^{N}S\bar{N^{c}}N$    
 with 
 VEVs of $H_{\nu}$ and $S$ as 
 $ \langle H_{\nu} \rangle \lll \langle H \rangle \ll \langle S
\rangle$,  
\begin{eqnarray}
\label{25} 
&\langle S \rangle \sim 1 \;\hbox{TeV},\;\;\;
 \langle H \rangle \sim 100 \;\hbox{GeV}, \;\;\;
\langle H_{\nu} \rangle \sim 10^{-0.5} \;\hbox{MeV}, 
\end{eqnarray}
neutrino mass is generated as 
\begin{equation}
m_\nu \bar{\nu^{c}_{L}}\nu_{L}\simeq \frac{y^{\nu2}\langle H_{\nu} \rangle ^{2}}
{y_{N}\langle S \rangle}\bar{\nu^{c}_{L}}\nu_{L}.
\end{equation}
This is so-called 
 type-I seesaw mechanism in a TeV-scale, where 
 coefficients 
 $y^\nu$ and $y^N$ are assumed to be 
 of order one. 
Notice that the 
 suitable scale of tiny neutrino mass of $O(0.1)$ eV is obtained. 

As for Higgs potential,
 it is given by
\begin{eqnarray}
V=&m^{2}|H|^{2}+m_{1}^{2}|H_{\nu}|^{2}-M^{2}|S|^{2}
 -m_{12}^{2}H^{\dagger}H_{\nu}
 -\lambda S^{3}-\mu SH^{\dagger}H_{\nu}\nonumber\\
     &+\frac{\lambda_{1}}{2}|H|^{4}+\frac{\lambda_{2}}{2}|H_{\nu}|^{4}+\lambda_{3}|H|^{2}|H_{\nu}|^{2}+\lambda_{4}|H^{\dagger}H_{\nu}|^{2}\nonumber\\
     &+\lambda_{S}|S|^4+
  \lambda_{H}|S|^{2}|H|^{2}+\lambda_{H_{\nu}}|S|^{2}|H_{\nu}|^{2}+h.c..
\end{eqnarray}
It should be noted that 
 interactions, such as $(H^\dagger H_{\nu})^{2}$, $H^\dagger H_{\nu}|H|^2$, 
 $H^\dagger H_{\nu}|H_{\nu}|^2$, $S^4$, $S^2|H|^2$, $S^2|H_{\nu}|^2$, etc, 
 are forbidden by $Z_{3}$-symmetry. 
The lepton number symmetry $U(1)_L$, 
 it is softly broken by both
 $\lambda$ and $\mu$ terms. 
Here we neglect mass term of $S^2$ ($\leq {\mathcal O}(1)$ TeV),
 for simplicity, 
 since the following analyses do not change as long as we do not 
 consider larger than ${\mathcal O}(1)$ TeV-mass of $S^2$ nor  
 CP violation in the Higgs sector. 
The mass term of 
 $m_{12}^2 H^\dagger H_{\nu}$, 
 which softly breaks $Z_{3}$-symmetry, 
 is introduced 
 to avoid domain-wall problem. 
Here 
 $|m_{12}^2|$ is assumed to be smaller than $|\mu \langle S \rangle |$ 
 (it means $m_{12}^2 \leq 10^{0.5}$ GeV in the following
 analyses),\footnote{
To obtain the desirable VEV-hierarchy of Eq.(\ref{218}) from 
 a stationary condition of Eq.(\ref{217}), 
 a condition $|m_{12}^{2}| < \frac{|\lambda_{3}
 \langle H_{\nu}\rangle \langle S \rangle^{2}|}{|\langle H \rangle |}$ 
 should be needed, which is automatically 
 satisfied under the condition of  
 $|m_{12}^2| < |\mu \langle S \rangle |$. 
}
 and this smallness (comparing to the weak scale) 
 against
 radiative corrections 
 is 
 guaranteed by the {\it softly breaking} $Z_3$-symmetry. 
Other 
 soft breaking $Z_3$-symmetry terms, such as $\mu' S|H|^{2}$, 
 are dropped, for simplicity, since  
 the following analyses are not changed as long as 
 $|\mu' \langle S \rangle| \leq |m^2|$. 

By denoting VEVs as\footnote{
Here we assume the VEVs for real. 
Case of complex VEVs can be analyzed similarly.
} 
\begin{equation}
\langle S \rangle=s,\;\;
\langle H \rangle=
\left(
\begin{array}{c}
h\\
0
\end{array}
\right),\;\;
\langle H_{\nu} \rangle=
\left(
\begin{array}{c}
h_{\nu}\\
0
\end{array}
\right), 
\end{equation}
stationary conditions
\begin{equation}
\frac{\partial V}{\partial s}=0,\;\;
\frac{\partial V}{\partial h}=0,\;\;
\frac{\partial V}{\partial h_{\nu}}=0,
\end{equation}
induce
 following equations, 
\begin{eqnarray}
\label{210}
&-2M^{2}s-6\lambda s^2-2\mu hh_{\nu}+4\lambda_{S}s^{3}+
2\lambda_{H}h^{2}s+2\lambda_{H_{\nu}}h_{\nu}^{2}s=0,
\\
\label{211}
&2m^{2}h-2\mu
 sh_{\nu}+2\lambda_{1}h^{3}+2\lambda_{3}hh_{\nu}^{2}+2\lambda_{4}hh_{\nu}^{2}+
 2\lambda_{H}hs^{2}-2m_{12}^{2}h_{\nu}=0,
\\
\label{212}
&2m_{1}^{2}h_{\nu}-2\mu sh+2\lambda_{2}h_{\nu}^{3}+
2\lambda_{3}h^{2}h_{\nu}+2\lambda_{4}h^{2}h_{\nu}+
2\lambda_{H_{\nu}}h_{\nu}s^{2}-2m_{12}^{2}h=0, 
\end{eqnarray}
respectively. 
Let us show conditions for the desirable vacuum, 
 $h_{\nu}\lll h \ll s$. 
The hierarchy of VEVs reduces Eq.(\ref{210}) to  
\begin{equation}
-2M^{2}s-6\lambda s^2+4\lambda_{S}s^{3}=0,
\end{equation}
which means 
\begin{equation}
s = \frac{M^{\prime}}{\sqrt{2\lambda_{S}}}
\end{equation}
where 
$M^{\prime}=\delta+\sqrt{M^{2}+\delta^{2}}$ and 
$\delta=\sqrt{\frac{9\lambda^{2}}{8\lambda_{S}}}$. 
The value of $\delta$ is small of order $10^{0.5}$ MeV
 as a scale of soft breaking of the lepton number symmetry,  
 which should be the same scale as 
 $\mu$ as shown later.

As for Eq.(\ref{211}),
 the hierarchy of VEVs reduces 
 the stationary condition as  
\begin{equation}
2(\lambda_{H}s^{2}+m^{2})h-2\mu sh_{\nu}+2\lambda_{1}h^{3}=0. 
\end{equation}
When 
\begin{equation}
\mu sh_{\nu} \ll \lambda_{1}h^{3}, 
\label{214}
\end{equation}
 we can neglect $\mu sh_{\nu}$-term, and 
 VEV of the SM-like Higgs becomes 
\begin{equation}
h=\frac{m^{\prime}}{\sqrt{\lambda_{1}}},
\end{equation}
where positive parameter $m^{\prime}$ is defined as 
\begin{equation}
\label{218}
m^{\prime 2}=-\lambda_{H}s^{2}-m^{2}.
\end{equation}
In the SM, 
 $m^{2}$ must be negative for so-called wine-bottle-type 
 potential. 
However, this model does not require negative mass squared 
 of $m^2 < 0$, 
 since
 the {\it effective negative mass squared} of 
 $m^{\prime 2} (=-\lambda_H s^2 -m^2) >0$ can be achieved with   
 a negative $\lambda_H$. 
The negative $\lambda_H$ does not break potential conditions of 
 bounded below, as long as 
 the value of $|\lambda_H|$ is smaller than 
 values of other four-point couplings, $\lambda$s, as 
 shown in Appendix A. 
So, one option is to take $\lambda_{1} \sim 1$ and
 $\lambda_H \sim -0.01$ which induces $m^{\prime 2}\sim
  {\mathcal O}(100^2)$ GeV$^2$, 
 and then  
 the suitable scale of 
 $\langle H \rangle \sim 100$ GeV is realized 
 through Eq.(\ref{218}) and 
 $\langle S \rangle \sim 1$ TeV. 

Finally, 
 the third stationary condition of Eq.(\ref{212}) 
 with the VEV-hierarchy becomes 
\begin{equation}
-2\mu sh+2\lambda_{H_{\nu}}h_{\nu}s^{2}-2m^{2}_{12}h=0. 
\label{217}
\end{equation}
Considering enough small $Z_3$-symmetry breaking 
 mass $m_{12}$,\footnote{This is the condition already shown in 
 the second previous 
 footnote. 
} 
 we obtain the VEV of $H_{\nu}$ as 
\begin{eqnarray}
h_{\nu}=\frac{\mu h}{\lambda_{H_{\nu}}s}. 
\label{218}
\end{eqnarray}
When we take $\mu\sim 10^{0.5}$ MeV, the desirable magnitude of 
 VEVs in 
 Eq.(\ref{25}) are reproduced, which is consistent with Eq.(\ref{214}). 
Since $\mu$ is soft breaking mass of the lepton number
 symmetry, its smallness is guaranteed against from 
 radiative corrections. 
Anyhow, notice that the 
 small magnitude of three-point mass parameter,
 $\mu$, plays a crucial role for 
 generating suitable magnitude of neutrino masses.

\subsection{Higgs mass spectra}

In a previous subsection, 
 we can obtain tiny VEV of extra Higgs doublet
 which is suitable for the magnitude of 
 neutrino masses through the seesaw mechanism. 
This is a nice feature, but
 do any light physical Higgs 
 particles appear due to the tiny VEV?  
Here, in this subsection, 
 we estimate mass spectra of physical Higgs 
 bosons. 

Denoting 
\begin{eqnarray}
&&S=s+{\rm Re}S+i{\rm Im}S, \nonumber \\
&&H=
\left(
\begin{array}{ccc}
h+{\rm Re}H^{0}+i{\rm Im}H^{0}\\
H^{-}
\end{array}
\right)
,\; \;  
H_{\nu}=
\left(
\begin{array}{ccc}
h_{\nu}+{\rm Re}H^{0}_{1}+i{\rm Im}H^{0}_{\nu}\\
H_{\nu}^{-}
\end{array}
\right),
\end{eqnarray}
a component of 
 Higgs mass matrix 
 is given by  
\begin{eqnarray}
M^{2}_{ij}=
\frac{1}{2}
  \frac{\partial^{2} V}{\partial v_i \partial v_j}
\end{eqnarray}
where
 $v_i$ means $s, h, h_{\nu}$. 
Here, the Higgs sector does not have CP violation, 
 so that $6\times6$ neutral Higgs mass matrix is given by 
\begin{eqnarray}
M^{2}_{Higgs}=
\left(
\begin{array}{cc}
M^{2}_{even}
&0\\
0
&M^{2}_{odd}
\end{array}
\right),
\end{eqnarray}
with 
\begin{eqnarray}
&M^{2}_{even}=
\small
\left(
\begin{array}{ccc}
- 3 \lambda s + 4\lambda_{S} s^{2}+\frac{\mu hh_{\nu}}{s}
&-2 \lambda_{H} h s - h_{\nu}\mu
&2 \lambda_{H_{\nu}} h_{\nu} s - h\mu
\\
-2 \lambda_{H} h s - h_{\nu}\mu
&2\lambda_{1} h^{2} +(m_{12}^{2}+s\mu)\frac{h_{\nu}}{h}
&2 \lambda_{3}hh_{\nu}+2\lambda_{4}hh_{\nu}-m_{12}^{2}-s\mu
\\
2 \lambda_{H_{\nu}} h_{\nu} s - h \mu
&2 \lambda_{3}hh_{\nu}+2\lambda_{4}hh_{\nu}-m_{12}^{2}-s\mu
&2\lambda_{2}h_{\nu}^{2}+(m_{12}^{2}+s\mu)\frac{h}{h_{\nu}}
\end{array}
\right),
\nonumber \\
&M^{2}_{odd}=
\small
\left(
\begin{array}{ccc}
9\lambda s+\frac{\mu hh_{\nu}}{s}
&-h_{\nu}\mu
&h\mu
\\
-h_{\nu}\mu
&(m_{12}^{2}+s\mu)\frac{h_{\nu}}{h}
&-m_{12}^{2}-s\mu
\\
h\mu
&-m_{12}^{2}-s\mu 
&(m_{12}^{2}+s\mu)\frac{h}{h_{\nu}}
\end{array}
\right). \nonumber 
\end{eqnarray}

For the CP even sector, 
 three physical (mass eigenstates) Higgs scalars
 are denoted as 
\begin{eqnarray}\small
\left(
\begin{array}{cc}
H_{S}
\\
h_{0}
\\
H_{0}  
\end{array}
\right)
=
\left(
\begin{array}{ccc}
c_{1}c_{2}& s_{1}c_{2}& s_{2}
\\
-s_{1}c_{3}+c_{1}s_{2}s_{3}& c_{1}c_{3}+s_{1}s_{2}s_{3}& -c_{2}s_{3}
\\
-s_{1}s_{3}-c_{1}s_{2}c_{3}& c_{1}s_{3}-s_{1}s_{2}c_{3}& c_{2}c_{3}
\end{array}
\right)^{\dagger}
\left(
\begin{array}{c}
{\rm Re} \; S
\\
{\rm Re} \; H
\\
{\rm Re} \; H_{\nu}
\end{array}
\right), 
\label{223}
\end{eqnarray}
where $c_{i}=\cos\alpha _{i},  
 s_{i}=\sin\alpha_{i}$ with 
\begin{eqnarray}
&\alpha _{1}=-\frac{2\lambda_{H}sh}{m^{2}_{H_{S}}-m^{2}_{h_{0}}}\;\;\;
\alpha_{2}=\frac{2\lambda_{H_{\nu}}sh_{\nu}-\mu h}
 {m^{2}_{H_{S}}-m^{2}_{H_{0}}}\;\;\;
\alpha_{3}=\frac{2
(\lambda_{3}+\lambda_{4})hh_{\nu}-m_{12}^{2}-\mu
s}{m^{2}_{H_{0}}-m^{2}_{h_{0}}}. 
\label{224}
\end{eqnarray}
The scalar 
 masses are given by 
\begin{eqnarray}
m^{2}_{H_{S}}=M^{2}+2\lambda_{S}s^{2},\;\;\;
m^{2}_{h_{0}}=2\lambda_{1}h^{2},\;\;\;
m^{2}_{H_{0}}=(m^{2}_{12}+\mu s)\frac{h}{h_{\nu}}. 
\end{eqnarray}
Under the condition of 
 $h_{\nu} \lll h \ll s$,  
 we know that 
 SM-like Higgs, $h_0$, is composed mainly of $H$ and  
 small components of $a_1(h/s) S + a_2(h_{\nu}/h) H_{\nu}$,
 where $a_i$s are order one coefficients. 
Similarly, 
 $H_0$
 is composed of $\sim H_{\nu} + a_3(h_{\nu}/h) H + a_4(h_{\nu}/s) S$ and 
 $H_S$
 is composed of $\sim H_S + a_5(h/s) H + a_6(h_{\nu}/s) H_{\nu}$.

Next, CP odd Higgs sector 
 has two Higgs pseudo-scalar, and one would-be NG boson
 which is absorbed into $Z$-boson. 
Two (mass eigenstates) pseudo-scalars and 
 would-be NG boson are denoted as 
\begin{eqnarray}
\left(
\begin{array}{c}
A_{S}
\\
\chi_{0}
\\
A_{0}
\end{array}
\right)
&=&
\left(
\begin{array}{ccc}
c^{\prime}_{2}& 0 & s^{\prime}_{2} \\
0 & c^{\prime}_{3} & -c^{\prime}_{2}s^{\prime}_{3} \\
-s^{\prime}_{2}c^{\prime}_{3}& +s^{\prime}_{3} & c^{\prime}_{2}c^{\prime}_{3}
\end{array}
\right)^{\dagger}
\left(
\begin{array}{c}
{\rm Im} S
\\ 
{\rm Im} H
\\
{\rm Im} H_{\nu}
\end{array}
\right)
\label{226}
\end{eqnarray}
where $c^{\prime}_{i}=\cos\beta_{i}$, 
 $s^{\prime}_{i}=\sin\beta_{i}$. 
Mixing angles are given by 
\begin{eqnarray}
&\tan\beta_{2}=\frac{\xi^{\prime} -\sqrt{\xi^{\prime
 2}+\eta^{\prime 2}+\eta^{\prime 2}_{1}}}{\xi
 +\sqrt{\xi^{2}+\eta^{2}+\eta^{2}_{1}}}, 
\;\; \tan\beta_{3}=\frac{h_{\nu}}{h},
\label{227}
\end{eqnarray} 
where $\xi=\frac{p}{q},\eta_{1}=\frac{h}{q},\eta_{2}=\frac{h_{\nu}}{q}$
 $\xi^{\prime}=\frac{p}{q^{\prime}},\eta^{\prime}_{1}=\frac{h}{q^{\prime}},\eta^{\prime}_{2}=\frac{h_{\nu}}{q^{\prime}},
 p=\frac{9\lambda s^{2}+\mu hh_{\nu}}{2\mu s}-\frac{m^{2}_{12}+\mu s}{2\mu h_{\nu}}h,
q^{2}=(p+\sqrt{p^{2}+h^{2}+h^{2}_{\nu}})^{2}+h^{2}+h^{2}_{\nu},
q^{\prime 2}=(p-\sqrt{p^{2}+h^{2}+h^{2}_{\nu}})^{2}+h^{2}+h^{2}_{\nu}$ .
Under the condition of 
 $h_{\nu} \lll h \ll s$,  
 two pseudo-Higgs bosons are given by 
\begin{eqnarray}
&A_{S}\sim \hbox{Im} S, 
&A_{0}\simeq \hbox{Im} H_{\nu},
\end{eqnarray}
which means 
 $A_{0}$ ($A_{S}$) is composed mainly of $H_{\nu}$ ($S$). 
These masses are given by 
\begin{eqnarray}
m^{2}_{A_{S}}=9\lambda s, \;\;\;
m^{2}_{A_{0}}= (m^{2}_{12}+\mu s)\frac{h}{h_{\nu}}.  
\end{eqnarray}
They are proportional to the {\it soft breaking} mass parameters,
 $\lambda$ and $\mu$, since they are 
 NG bosons of global $U(1)$ symmetries. 
It should be noticed that $A_0$ has the same as $H_0$. 
Supposing $\lambda=0$ and $m_{12}^2=0$, 
 Lagrangian has an accidental global symmetry, 
\begin{eqnarray}
H \to e^{-i\theta_1}H, \;\;\;
H_{\nu} \to e^{i\theta_1}H_{\nu}, \;\;\; 
S \to e^{-i2\theta_1}S. 
\end{eqnarray}
Then massless NG boson appears after the symmetry breaking caused by 
 VEVs of Higgs fields. 
Similarly, 
 when $\mu=0$ and $m_{12}^2=0$, 
 there exists 
 a global symmetry, 
\begin{eqnarray}
H \to e^{i\theta_2}H, \;\;\;
H_{\nu} \to e^{-i\theta_2}H_{\nu}, 
\end{eqnarray}
which induces 
 a 
 massless NG boson after the symmetry breaking. 
These mass parameters $\lambda$ and $\mu$ also break lepton number 
 symmetry, so that 
 the pseudo-scalars can be regarded as so-called 
 ``Majoron''. 
But they are heavier than the SM-like Higgs as long as 
 $\lambda \geq {\mathcal O}(10)$ GeV. 



As for the charged Higgs sector, 
 would-be NG boson, $\chi^+$, and physical state, $h^+$,  are given by 
\begin{eqnarray}
&\chi^{+}=\cos\beta_{3} H^{+}-\sin\beta_{3} H^{+}_{\nu}, \;\;\;
h^{+}=\sin\beta_{3} H^{+}+\cos\beta_{3} H^{+}_{\nu}.  
\end{eqnarray}
The charged Higgs mass is given by 
\begin{eqnarray}
&m^{2}_{h^{+}}=-\lambda_{4}(h^{2}+h_{\nu}^{2})+\frac{2(m_{12}^{2}+\mu
 s)}{\sin 2\beta_{3}}. 
\label{ch}
\end{eqnarray}
Note that 
 the second term is almost same as 
 the masses of $H_0$ and $A_0$ due to 
 $h/h_\nu \gg 1$, and 
 the mass difference between $m_{h^\pm}^2$ and $m_{H_0}^2$, $m_{A_0}^2$ 
 is just a weak scale squared from the first term. 
Charged Higgs 
 plays crucial roles of phenomenology, such as 
 lepton flavor violating processes. 
We show some phenomenology induced from the 
 charged Higgs boson in the next section. 


Before ending of this section, 
 we comment on
 limits of $h/s\rightarrow 0$ and $h_{\nu}/h\rightarrow 0$.     
In the limits, 
 the SM-like Higgs is just $H$ and its physical state 
 is physical neutral Higgs, $h_0$, and 
 an imaginary part and charged components are
 absorbed by $Z$ and $W^\pm$. 
As for a singlet Higgs, $S$, and an extra doublet Higgs, $H_{\nu}$,
 they are origins of other physical Higgs particles, 
 $H_S, A_S$, and $H_0, A_0, h^\pm$, 
 respectively.  
Notice that these approximations are 
 justified up to ratios of VEVs. 

\section{Phenomenology}

This section is devoted to some phenomenology of our model. 
We show decay of Higgs bosons, 
 lepton flavor violation process, 
 $\rho$ parameter, 
 neutrinoless double beta decay, 
 and so on. 

\subsection{Decay of charged-Higgs boson}

Since the charged Higgs mass is given by Eq.(\ref{ch}), 
 it becomes smaller or larger than masses of 
 $H_0, A_0$ depending on a sign of $\lambda_4$. 
There is also a possibility that 
 the charged Higgs mass is smaller or larger 
 than masses of right-handed neutrinos. 
Thus, 
 a dominant process of charged Higgs decay 
 depends on the mass spectra of them. 
We will show four cases according to 
 $m_{h^\pm}<m_{H_0,A_0}$ or $m_{h^\pm}>m_{H_0,A_0}$
 and $m_{h^\pm}<m_N$ or $m_{h^\pm}>m_N$,
 as follows. 
A important point is that the decay of charged Higgs to quarks and 
 charged leptons are strongly suppressed due to the 
 absence of direct couplings among them, which 
 is one of the different points from the usual 
 two Higgs double models.

\subsubsection{$m_{h^\pm}<m_{H_0,A_0}, \; m_N$}

At first, let us show the case of 
 {$m_{h^\pm}<m_{H_0,A_0}, \; m_N$}. 
In this case, only possible charged Higgs decay modes are 
 to quarks and
 charged leptons through the Yukawa interactions of
 Eq.(\ref{22}).  
Since the charged Higgs is mainly composed by $H_{\nu}$, 
 its coupling with 
 quarks and charged leptons are
 always suppressed by $\sim h_{\nu}/h$. 
Thus, this case tends to induce long life time 
 of charged Higgs comparing to cases of 
 other mass spectra. 
The effective Yukawa interactions between $h^+$ and 
 quarks and charged leptons are given by
\begin{eqnarray}
L_{yukawa}=
(y^{d}_{ij}h^{+}\bar{u}_{L_{i}}d_{R_{j}}+
y^{u}_{ij}h^{+}\bar{d}_{L_{i}}u_{R_{j}}+
y^{l}_{ij}h^{+}\bar{\nu_{i}}l_{R_{j}})\sin\beta_{3}+
y^{N}_{ij}h^{+}\bar{l}_{L_{i}}N_{j}\cos\beta_{3}
\label{eff}
\end{eqnarray}
Then,
 the total decay width is given by 
\begin{eqnarray}
\Gamma_{tot}
&=&\Gamma(h^{+} \to u_{L_{i}}\bar{d}_{R_{j}})
+\Gamma(h^{+} \to \bar{d}_{L_{i}}{u}_{R_{j}})
+\Gamma(h^{+} \to \nu_{i}\bar{l}_{R_{j}}) \nonumber \\
&=&\sum_{i,j}\frac{3m_{h^{+}}}{16\pi}
\sin^2\beta_{3}
(\big| y^{d}_{ij}\big|^{2}+\big| y^{u}_{ij}\big|^{2}+\frac{1}{6}\big|
y^{l}_{ij}\big|^{2}). 
\label{336}
\end{eqnarray}
It means 
 the charged Higgs almost 
 decays to right-handed top and left-handed bottom quarks due
 to the large Yukawa coupling. 
Using 
 $\sin\beta_3\simeq h_\nu/h$, 
 the life time of charged Higgs is given by
\begin{equation}
\tau (h^\pm)\sim 10^{-16} s.  
\end{equation}
It means the charged Higgs 
 propagates 
 a very short distance which can not be detected  
 in the detector of collider experiments.

\subsubsection{$m_N<m_{h^\pm}<m_{H_0,A_0}$} 

Next, we show the case of 
 {$m_N< m_{h^\pm}<m_{H_0,A_0}$. 
In this case, the charged Higgs can decay 
 to (left-handed) charged leptons and 
 right-handed neutrinos through the 
 Yukawa
 interaction of $y^{\nu}\bar{L}H_{\nu}N$ in Eq.(\ref{eff}), 
 which has no suppression factor because of $\cos\beta_{3}\simeq 1$. 
Then, the decay width is given by 
\begin{eqnarray}
\Gamma(h^{+} \to N_{i}l_{L_{j}})
=\frac{m_{h^{+}}}{32\pi}\big| y^{\nu}_{ij}\big|^{2}\Big(1-\frac{m^{2}_{N_{i}}}{m^{2}_{h^{+}}}\Big).
\end{eqnarray}
Remind that in 
 the usual two Higgs doublet model, 
 the charged Higgs mainly decay to 
 the heavy quarks. 
While, in our model with this mass spectrum, 
 the charged Higgs mainly decays to 
 charged leptons and right-handed neutrinos.  
When the right-handed neutrinos are missing 
 in the collider experiments, 
 this is a single charged lepton event 
 with missing transverse momentum, which can 
 be clearly detected in the detector. 
Especially, 
 the case that $y^\nu$ of the first and second generations 
 are larger than that of the third generation is interesting, which  
 induces electron and muon events in 
 collider experiments, and they 
 can be clearly detected.  
This situation can be consistent with 
 any neutrino mass hierarchies through the seesaw 
 mechanism with a suitable mass hierarchy 
 of right-handed neutrinos. 
Notice that 
 this 
 situation can not be realized in 
 case of Dirac neutrino scenario\cite{{Ma}, {Nandi}, {Davidson:2009ha}, {Logan:2010ag}}.

\subsubsection{$m_{H_0,A_0}<m_{h^\pm}<m_N$}

Next is devoted to 
 the case of {$m_{H_0,A_0}<m_{h^\pm}<m_N$}. 
The dominant charged Higgs decay mode is 
 $h^\pm \to W^\pm H_0, A_0$ through the gauge interaction. 
The decay width is given by 
\begin{eqnarray}
\Gamma(h^{+} \to W^+ H_0, A_0)
&\simeq&
\frac{g^{2}_{2}m^{3}_{h^{+}}}{16\pi
m^{2}_{W}}\Big(1-\frac{m^{2}_{H_{0},A_{0}}}{m^{2}_{h^{+}}}\Big)^{3}, 
\end{eqnarray}
where $g_2$ is the gauge coupling of 
 weak interaction. 
Notice that the decay processes to quarks and 
 charged leptons are strongly suppressed due to 
 the suppression factor, $\sin\beta_3\simeq h_{\nu}/h$, as  
 Eq.(\ref{336}). 
This is one of the different points from the usual 
 two Higgs double models   
 where the main decay mode is heavy quarks.

\subsubsection{$m_{h^\pm}>m_{H_0,A_0}, \; m_N$}

Finally, 
 let us show the case of 
 $m_{h^\pm}>m_{H_0,A_0}, \; m_N$. 
In this case, the charged Higgs $h^+$ 
 can decay both 
 to $W^+ H_0, A_0$ and 
 $N_{i} l_{L_{j}}$. 
They have no suppression factor from $h_\nu/h$,   
 so that 
 each decay width is given by 
\begin{eqnarray}
&&\Gamma(h^{+} \to W^+ H_0, A_0)
\simeq
\frac{g^{2}_{2}m_{h^{+}}^3}
{16\pi
m^{2}_{W}}\Big(1-\frac{m^{2}_{H_{0},A_{0}}}{m^{2}_{h^{+}}}\Big)^{3}, 
\\
&&\Gamma(h^{+} \to N_{i}l_{L_{j}})
=\frac{m_{h^{+}}}{32\pi}\big| y^{\nu}_{ij}\big|^{2}\Big(1-\frac{m^{2}_{N_{i}}}{m^{2}_{h^{+}}}\Big).
\end{eqnarray}
The dominant decay mode depends on the magnitude of 
 $|y^\nu|$ and degeneracy factor of $m_h^\pm$  
 and $m_{H_0, A_0}$, $m_N$. 
Thus, the main mode can not be determined until 
 a concrete mass spectrum is fixed. 
One interesting example is a case of 
 $m_{h^\pm}\geq m_{H_0,A_0}>m_N$. 
Taking 
 $y^\nu \sim 1$ 
 for the heaviest neutrino 
 and degenerate right-handed Majorana masses, 
 and also considering 
 mass hierarchy 
 of active neutrinos,  
 inverted (normal) hierarchy, IH (NH),   
 induces single left-handed muon (tau) event 
 with missing transverse momentum as a dominant
 decay mode.\footnote{
Notice that 
 this is quite different point from 
 the usual two Higgs doublet models. 
}

\subsection{$\rho$ parameter}

Next, let us estimate 
 charged Higgs contribution to 
 $\rho$ parameter, which 
 is almost same as usual two Higgs doublet models\cite{{2hdm}} 
 due to the small mixings between 
 the singlet Higgs $S$ and Higgs doublets $H,H_{\nu}$. 
It is estimated as 
\begin{eqnarray}
\delta\rho
=\sqrt{2}G_{F}\frac{1}{(4\pi)^{2}}
\Big[F_{\Delta}(m^{2}_{A_{0}},m^{2}_{h^{\pm}})
-s^{2}_{\alpha-\beta}\big[F_{\Delta}(m^{2}_{h_{0}},m^{2}_{A_{0}})-F_{\Delta}(m^{2}_{h_{0}},m^{2}_{h^{\pm}})\big]\nonumber\\
-c^{2}_{\alpha-\beta}\big[F_{\Delta}(m^{2}_{H_{0}},m^{2}_{A_{0}})-F_{\Delta}(m^{2}_{H_{0}},m^{2}_{h^{\pm}})\big]
\Big],
\end{eqnarray}
where $c_{\alpha-\beta}=\cos(\alpha_3-\beta_3)$,
 $s_{\alpha-\beta}=\sin(\alpha_3-\beta_3)$, and 
\begin{eqnarray}
F_{\Delta}(x,y)=\frac{1}{2}(x+y)-\frac{xy}{x-y}\ln\frac{x}{y}.
\end{eqnarray}
The $\alpha_{3}$ represents (almost) mixing angle  
 between $h_0$ and $H_0$, and  
 $h_0$ is almost 
 SM-like Higgs since 
 $c_{\alpha-\beta}\simeq 1$.  
The mass spectrum shows 
 $h^\pm$ and $A_0$ are degenerate
 in TeV-scale as $\Big|\frac{m^{2}_{h^{+}}-m^{2}_{A_{0}}}{m^{2}_{h^{+}}}\Big|
=\Big|\frac{h^{2}}{m^{2}_{h^{+}}}\Big|\sim0.01$.
Thus, $\delta \rho$ is estimated as 
\begin{eqnarray}
\delta\rho_{2HDM}
&\simeq &\frac{2\sqrt{2}G_{F}}{(4\pi)^{2}}F_{\Delta}(m^{2}_{A_{0}},m^{2}_{h^{\pm}})
\simeq\frac{\sqrt{2}G_{F}}{3(4\pi)^{2}}\frac{\lambda_{4,}h^{2}}{m_{h^{+}}}
\sim 10^{-7} ,
\end{eqnarray}
which means the correction to $\rho$ parameter
 is negligible in our model.

\subsection{Decay of $h_{0}$}

Here we show a decay of SM-like Higgs $h_0$,
 which has tiny coupling with neutrinos due to the 
 small mixing $\sim \sin\alpha_3$.
In our setup, Higgs mass spectrum is given by 
 $m_{h_{0}}<2m_{h^{+},H_{0},A_{0},H_{S},A_{S}}$,
 so that the SM-like Higgs 
 $h_0$ decay to quarks and charged leptons
 through the usual Yukawa interactions.\footnote{  
We comment on the case of 
 $m_{h_{0}}>2m_{h^{+},H_{0},A_{0},H_{S},A_{S}}$,  
 which is possible by changing the hierarchy of VEVs  
 although it is out of our aim.  
Anyway, in this case, 
 decay channels of 
 $h_{0} \to h^{+}h^{-},2H_{0},2A_{0},2H_{S},2A_{S}$ 
 open through 
 the 
 mixings among three Higgs fields, ($H,H_{\nu},S$).  
Their decay widths are given by 
\begin{eqnarray}
&&\Gamma(h_{0} \to
 h^{+}h^{-})=\frac{\lambda^{2}_{3}m_{h_{0}}}{16\pi\lambda_{1}}\sqrt{1-\frac{4m^{2}_{h^{\pm}}}{m^{2}_{h_{0}}}},\;\;\;\;\;
 \Gamma(h_{0} \to H_{S}H_{S},A_{S},A_{S})=\frac{\lambda^{2}_{H}m_{h_{0}}}{32\pi\lambda_{1}}\sqrt{1-\frac{4m^{2}_{H_{S},A_{S}}}{m^{2}_{h_{0}}}},\nonumber \\
&&\Gamma(h_{0} \to H_{0}H_{0})=\Gamma(h_{0}\to A_{0}A_{0})=\frac{(\lambda_{3}+\lambda_{4})^{2}m_{h_{0}}}{32\pi\lambda_{1}}\sqrt{1-\frac{4m^{2}_{H_{0},A_{0}}}{m^{2}_{h_{0}}}},
\nonumber 
\end{eqnarray}
respectively. 
}
And the main mode is of cause top and bottom quarks
 due to the large Yukawa coupling.

As for a process of $h_0\to \gamma\gamma$, which has 
 tiny background, 
 the decay width is modified due to the charged Higgs 
 loop contribution, which is given by  
\begin{eqnarray}
\Gamma(h_{0}\to\gamma\gamma)
&=&\Gamma^{SM}(h_{0}\to\gamma\gamma)\big[1-\lambda_{3}\delta\big(\frac{100GeV}{M_{h^{+}}}\big)^{2}\big]^{2},
\nonumber \\
&\simeq&0.997\times\Gamma^{SM}(h_{0}\to\gamma\gamma). 
\end{eqnarray}
Where $\delta=0.16$ for 1 TeV charged Higgs\cite{Davidson:2009ha}. 
Thus, in our model with TeV-scale mass of the charged Higgs,
 the modification of $h_0\to \gamma\gamma$ is tiny, 
 less than ${\mathcal O}(1)$ \%.

\subsection{Lepton flavor violation \& anomalous magnetic moment}

Let us estimate lepton flavor violating process induced from 
 charged Higgs boson 1-loop diagrams. 
Remind that Yukawa interactions of neutrinos in Eq.(\ref{22}) 
 are given by   
\begin{equation}
\frac{1}{2}y^{N}_{ij}S\bar{N^{c}_{i}}N_{j}
 +y^\nu_{ij}\bar{N^{c}_{i}}(\nu_{j}H^{0}_{\nu}-l_{jL}H^{+}_{\nu})
 +{\rm h.c.}. 
\end{equation}
Here we assume 
\begin{equation}
y^{N}_{ij}=\frac{M_{M}}{\langle S \rangle} \delta_{ij}, 
\end{equation}
for simplicity. 
$M_M$ is a mass parameter of order TeV-scale. 
Then, the 
 mass matrix of the light neutrinos become 
\begin{equation}
M_{\nu}=\frac{h_{\nu}^2}{M_{M}}\sum_{j}y^\nu_{ij}y^\nu_{ij}. 
\end{equation}
Noting $U y^\nu y^{\nu T}
 U^T={\rm diag.}(4y^{\nu2}_1,4y^{\nu2}_2,4y^{\nu2}_3)$  
 where $U$ is the MNS matrix, 
 Yukawa coupling 
 $y^\nu_{ij}=2y^\nu_i \delta_{ik} (U^T)_{kj}$
 is given by 
\begin{eqnarray}
y^\nu_{ij}\simeq
\left(
\begin{array}{ccc}
  \sqrt{3}cy^{\nu}_{1}&   -\frac{1}{\sqrt{2}}(1+\sqrt{3}s)y^{\nu}_{1}&    \frac{1}{\sqrt{2}}(1-\sqrt{3}s)y^{\nu}_{1}\\
  cy^{\nu}_{2}&   \frac{1}{\sqrt{2}}(\sqrt{3}-s)y^{\nu}_{2}&   -\frac{1}{\sqrt{2}}(\sqrt{3}+s)y^{\nu}_{2}\\
  2sy^{\nu}_{3}&   \sqrt{2}cy^{\nu}_{3}&   \sqrt{2}cy^{\nu}_{3}
  \end{array}
\right). 
\end{eqnarray}
Where we note $s=\sin\theta_{13}, c=\cos\theta_{13}$, 
 and take $\theta_{12}=\pi/6, \theta_{23}=\pi/4$. 

A branching ration of 
 $l_{i} \to l_{j} \gamma$ to 
 $l_{i} \to l_{j} \nu_{i} \bar{\nu_{j}}$
 is given by\cite{Ma:2001mr} 
\begin{eqnarray}
R(l_{i} \to l_{j} \gamma)=\frac{192 \pi^3
 \alpha}{G^{2}_{F}m^{4}_{l_{i}}}\left| 
\sum_{k}\frac{y^\nu_{kl_{i}}y^\nu_{kl_{j}}}
{48(4\pi)^{2}}\frac{m^{2}_{l_{i}}}{m^{2}_{h^{+}}} 
 \right|^{2}.
\end{eqnarray}
where $\alpha$ is the fine structure constant 
 $\alpha=e^{2}/4\pi$ and 
 $m_{l_i}$ is the $i$-th generation
 charged lepton mass. 
For example, 
 by using $Br(\mu \to e \nu \bar{\nu})\simeq 1$, 
 the branching ration $\mu \to e \gamma$ is given by 
\begin{eqnarray}
Br(\mu \to e \gamma)
&\simeq&
\frac{\alpha M^{2}_{M}}{98304\pi G^{2}_{F}m^{4}_{h^{+}}h^{4}_{\nu}}\cos^{2}\theta
\bigl(\sqrt{3}\delta m_{12}+\sin\theta(3\delta m_{12}+
 4\delta m_{23})\bigr)^{2}, \\
&=&
\frac{\pi\alpha}{384G^{2}_{F}}\Big(\frac{\alpha_{{\nu}_{i}}}{m^{2}_{h^{+}}}\Big)^{2}\frac{\cos^{2}\theta}{m^{2}_{\nu_{i}}}
\bigl(\sqrt{3}\delta m_{12}+\sin\theta(3\delta m_{12}+4\delta m_{23})\bigr)^{2},
\end{eqnarray}
where 
 $\alpha_{\nu_{i}}\equiv\frac{y^{2}_{\nu_{i}}}{4\pi}$, 
 $\theta=\theta_{13}$, 
 $m_{{\nu}_{i}}$ is the lightest neutrino mass 
  (which means $m_\nu=m_{\nu_{1}}$ in the NH and 
  $m_\nu=m_{\nu_{3}}$ in the IH as will be shown in
  Eqs.(\ref{NH}) and (\ref{IH})), and  
\begin{eqnarray}
\delta m_{ij}\equiv m_{j}-m_{i}, \;\;\;\;\;\;
\Delta m^{2}_{ij}\equiv m^{2}_{j}-m^{2}_{i}. 
\end{eqnarray}
Generation dependence of neutrino mass, $m_i$, depends on 
 neutrino mass hierarchy, 
 NH or IH. 
By using neutrino oscillation experimental data\cite{Strumia:2006db}, 
\begin{eqnarray}
\Delta m^{2}_{\odot}\simeq7.6\times10^{-5}\ {\rm eV}^{2}, \;\;\;\;\;
\Delta m^{2}_{atm}\simeq2.4\times10^{-3}\ {\rm eV}^{2}, 
\end{eqnarray}
and the lightest neutrino mass, $m_{\nu_i}$, 
 NH shows 
\begin{eqnarray}
m_{1}&=&m_{\nu}, \;\;\;\;\;
m_{2}=\sqrt{m^{2}_{\nu}+\Delta m^{2}_{\odot}}, \;\;\;\;\; 
m_{3}=\sqrt{m^{2}_{\nu}+\Delta m^{2}_{\odot}+\Delta m^{2}_{atm}}, 
\label{NH}
\end{eqnarray}
and IH shows 
\begin{eqnarray}
m_{1}&=&\sqrt{m^{2}_{\nu}-\Delta m^{2}_{\odot}+\Delta m^{2}_{atm}}, \;\;\;\;\;
m_{2}=\sqrt{m^{2}_{\nu}+\Delta m^{2}_{atm}}, \;\;\;\;\;
m_{3}=m_{\nu},
\label{IH} 
\end{eqnarray}
respectively. 
In case of degenerate neutrino masses, 
 both NH and IH become 
\begin{eqnarray}
Br(\mu\to e\gamma)\to 
\frac{\pi\alpha}{1536G^{2}_{F}}\Big(\frac{\alpha_{\nu_{i}}}{m^{2}_{h^{+}}}\Big)^{2}\frac{\cos^{2}\theta}{m^{4}_{\nu}}
\bigl(\sqrt{3}\Delta m^{2}_{12}+\sin\theta(3\Delta m^{2}_{12}+4\Delta
m^{2}_{23})\bigr)^{2}. 
\label{359}
\end{eqnarray}
It means 
 the branching ratio decreases as $m^{-4}_{\nu_{i}}$ in 
 the degenerate hierarchy region 
 which can be shown in Figures 2.

As for processes of 
 $\tau\to\mu\gamma$ and $\tau\to e\gamma$, 
 they are given by 
\begin{eqnarray}
R(\tau\to\mu\gamma)&=&\frac{192\pi^{3}\alpha}{G^{2}_{F}m^{4}_{\tau}}\Big|\sum_{k}\frac{y^{\nu}_{k\tau}y^{\nu}_{k\mu}}{48(4\pi)^{2}}\frac{m^{2}_{\tau}}{m^{2}_{h^{+}}}\Big|^{2}, \\
R(\tau\to
 e\gamma)&=&\frac{192\pi^{3}\alpha}{G^{2}_{F}m^{4}_{\tau}}\Big|\sum_{k}\frac{y^{\nu}_{k\tau}y^{\nu}_{ke}}{48(4\pi)^{2}}\frac{m^{2}_{\tau}}{m^{2}_{h^{+}}}\Big|^{2}, 
\end{eqnarray}
where 
\begin{eqnarray}
\sum_{k}y^{\nu}_{k\tau}y^{\nu}_{k\mu}
&=&\frac{M_{M}}{8h^{2}_{\nu}}\Big((\delta m_{12}-4\delta
 m_{23})-\sin^{2}\theta (3\delta m_{12}+4\delta m_{23})  \Big), \\
\sum_{k}y^{\nu}_{k\tau}y^{\nu}_{ke}
&=&-\frac{M_{M}\cos\theta}{4\sqrt{2}h^{2}_{\nu}}\Big(\delta m_{12}
 -\sin\theta(3\delta m_{12}+4\delta m_{23}) \Big). 
\end{eqnarray}
Thus, branching ratios are calculated as 
\begin{eqnarray}
Br(\tau\to\mu\gamma)
&=&Br(\tau\to\mu\nu\bar{\nu})
\frac{\pi\alpha}{768G^{2}_{F}}\Big(
\frac{\alpha_{\nu_{i}}}{m^{2}_{h^{+}}}\Big)^{2}\Big(\frac{(\delta
 m_{12}+4\delta m_{23})-\sin^{2}\theta_{}(3\delta m_{12}+4\delta
 m_{23})}{m_{\nu_{i}}}\Big)^{2}, \nonumber \\ 
&& \hspace*{-2.5cm}\to Br(\tau\to\mu\nu\bar{\nu}) 
\frac{\pi\alpha}{3072G^{2}_{F}}\Big(\frac{\alpha_{\nu_{i}}}{m^{2}_{h^{+}}}\Big)^{2}\Big(\frac{(\Delta
m^{2}_{12}+4\Delta m^{2}_{23})-\sin^{2}\theta_{}(3\Delta
m^{2}_{12}+4\Delta m^{2}_{23})}{m^{2}_{\nu_{i}}}\Big)^{2}, 
\end{eqnarray}
\begin{eqnarray}
Br(\tau\to e\gamma)
&=&Br(\tau\to e\nu\bar{\nu})
\frac{\pi\alpha}{384G^{2}_{F}}\Big(\frac{\alpha_{{\nu}_{i}}}{m^{2}_{h^{+}}}\Big)^{2}\frac{\cos^{2}\theta}{m^{2}_{\nu_{i}}}
\bigl(\sqrt{3}\delta m_{12}-\sin\theta 
 (3\delta m_{12}+4\delta m_{23})\bigr)^{2} \nonumber \\
&& \hspace*{-2.5cm}\to Br(\tau\to e\nu\bar{\nu}) 
\frac{\pi\alpha}{1536G^{2}_{F}}\Big(\frac{\alpha_{\nu_{i}}}{m^{2}_{h^{+}}}\Big)^{2}\frac{\cos^{2}\theta}{m^{4}_{\nu}}
\bigl(\sqrt{3}\Delta m^{2}_{12}+\sin\theta(3\Delta m^{2}_{12}+4\Delta
m^{2}_{23})\bigr)^{2}, 
\label{365}
\end{eqnarray}
respectively. 
Where the second line in each equation
 is degenerate neutrino mass limit, 
 and we use 
 $Br(\tau\to\mu\nu\bar{\nu})\simeq 0.17$ and   
 $Br(\tau\to e\nu\bar{\nu})\simeq 0.18$\cite{PDG} in 
 the following numerical calculations.

Figures 1 show 
 $\theta$ dependence of 
 branching ratios of 
 $\mu\to e\gamma$ (red line),
 $\tau\to e\gamma$ (blue line),
 and $\tau\to \mu\gamma$ (green line) 
 with 
 $\alpha_{\nu}=1/4\pi$, $m_{\nu}=0.1$ eV, and $m_{h^{+}}=1$ TeV. 
Dashed lines correspond to experimental bound\cite{{Brooks:1999pu},{Hayasaka:2009zz}}. 
NH (IH) has decreasing point at $\theta\simeq 0.014$ 
 ($\theta_{}\simeq0.013$) in $Br(\tau\to e\gamma)$ 
 ($Br(\mu\to e\gamma)$), 
 which can be understood from a cancellation in Eq.(\ref{365}) 
 (Eq.(\ref{359})). 

Figures 2 are 
 the lightest mass $m_{\nu}$ dependence of 
 branching ratios of $\mu\to e\gamma$, $\tau\to e\gamma$,
 and $\tau\to \mu\gamma$ with 
 $\alpha_{\nu}=1/4\pi$ and $m_{h^{+}}=1$ TeV. 
Red line shows 
 $\theta_{}=0$, 
 green line 
 $\theta_{}=0.001$, 
 blue line 
 $\theta_{}=0.01$, and 
 purple line 
 $\theta_{}=0.1$. 
Dashed line corresponds to each experimental bound. 
The decreasing point in IH is also calculated from  
 a cancellation in Eq.(\ref{359}).
When we take $\theta_{}= 0.013$ in $Br(\mu\to e\gamma)$ with IH, 
 a decreasing point emerges at $m_{\nu}\simeq0.1$ eV, 
 which is consistent with Figures 1. 

Figures 1 and 2 show that a wide parameter region can be reached 
 by 
 the MEG experiment 
 which has a sensitivity of order $10^{-13}$\cite{{MEG}}. 

\begin{figure}[htbp]
\begin{center}
\includegraphics[width=7cm, bb=0 0 240 173]{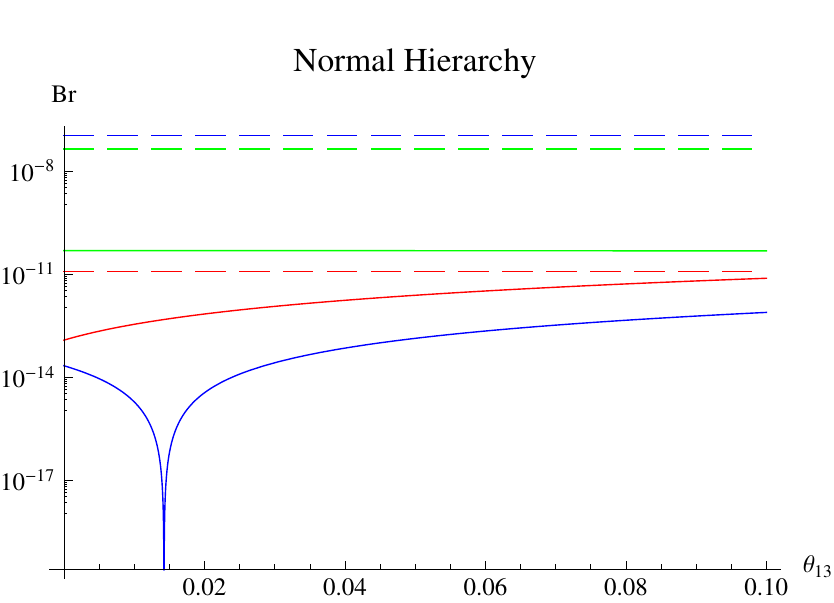}\;
\includegraphics[width=7cm, bb=0 0 240 173]{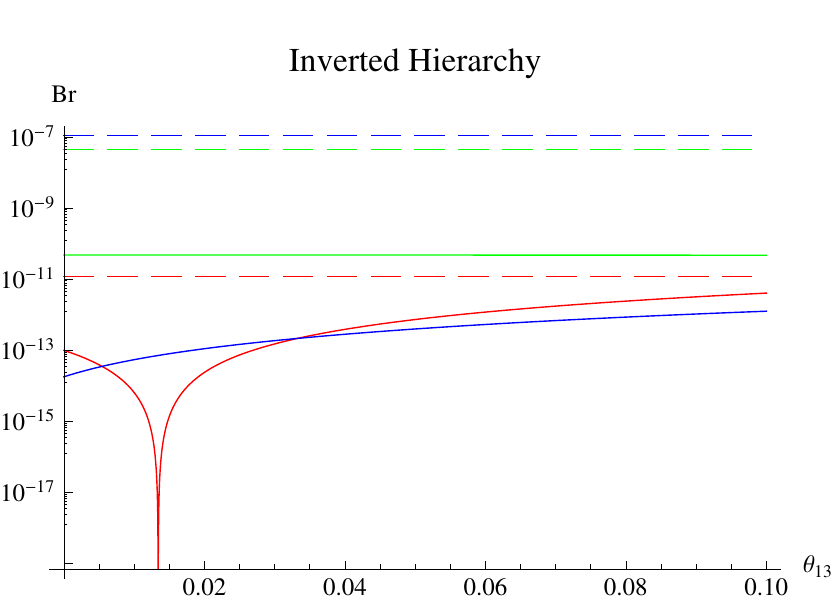}
\caption{
$\theta$ dependence of 
 branching ratios of 
 $\mu\to e\gamma$ (red line),
 $\tau\to e\gamma$ (blue line),
 and $\tau\to \mu\gamma$ (green line) 
 with 
 $\alpha_{\nu}=1/4\pi$, $m_{\nu}=0.1$ eV, and $m_{h^{+}}=1$ TeV. 
Dashed lines correspond to experimental bound. 
}
\label{fig1}
\end{center}
\end{figure}

\begin{figure}[htbp]
\begin{center}
\includegraphics[width=7cm, bb=0 0 240 173]{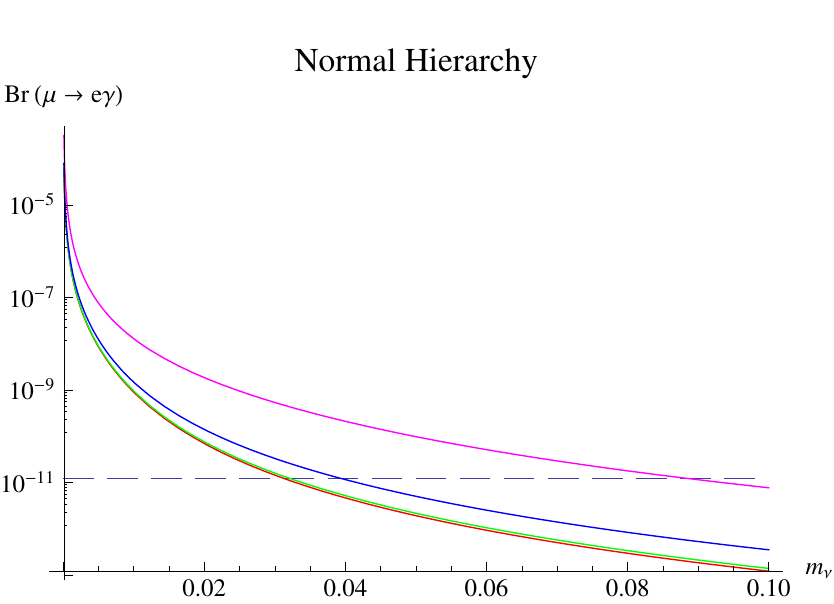}
\includegraphics[width=7cm, bb=0 0 240 173]{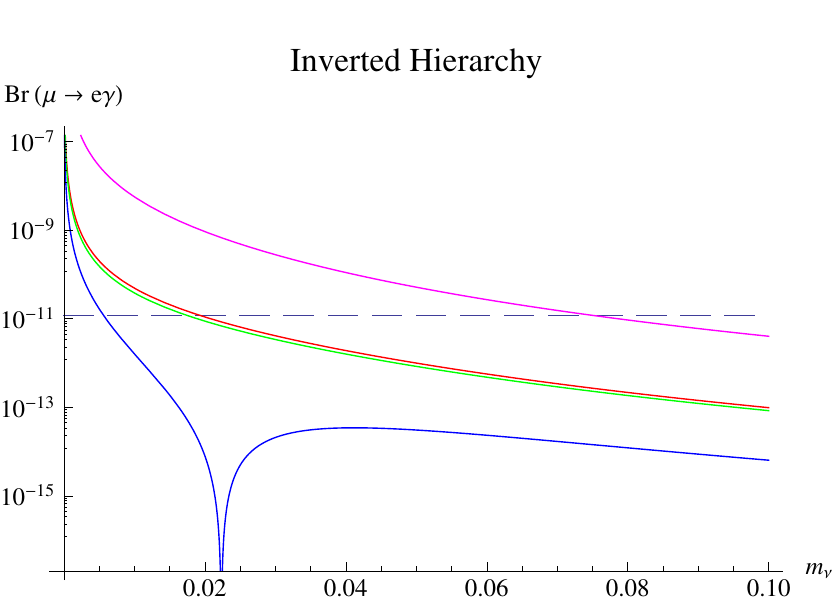}
\includegraphics[width=7cm, bb=0 0 240 173]{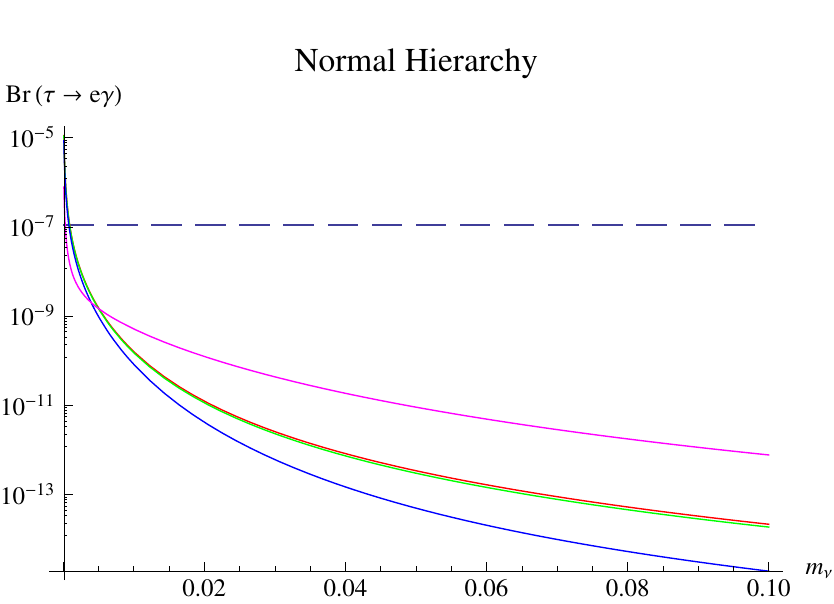}
\includegraphics[width=7cm, bb=0 0 240 173]{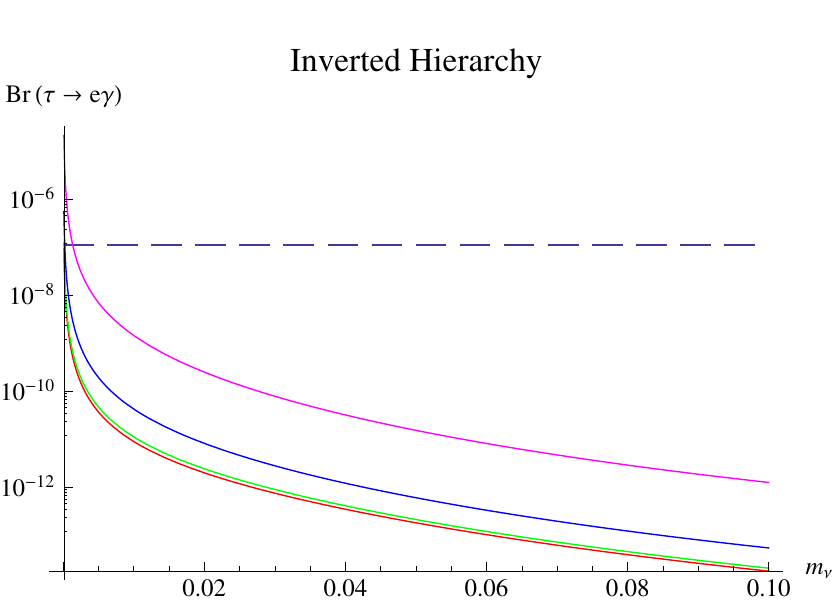}
\includegraphics[width=7cm, bb=0 0 240 173]{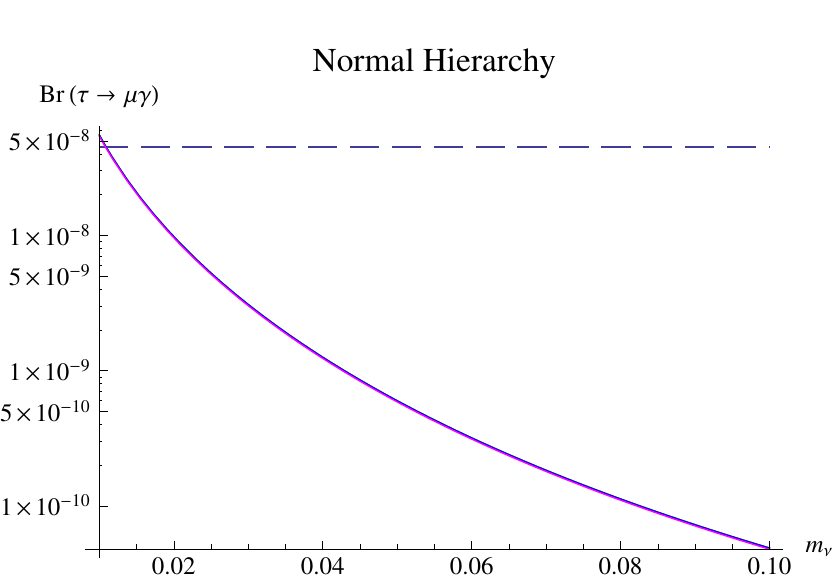}
\includegraphics[width=7cm, bb=0 0 240 173]{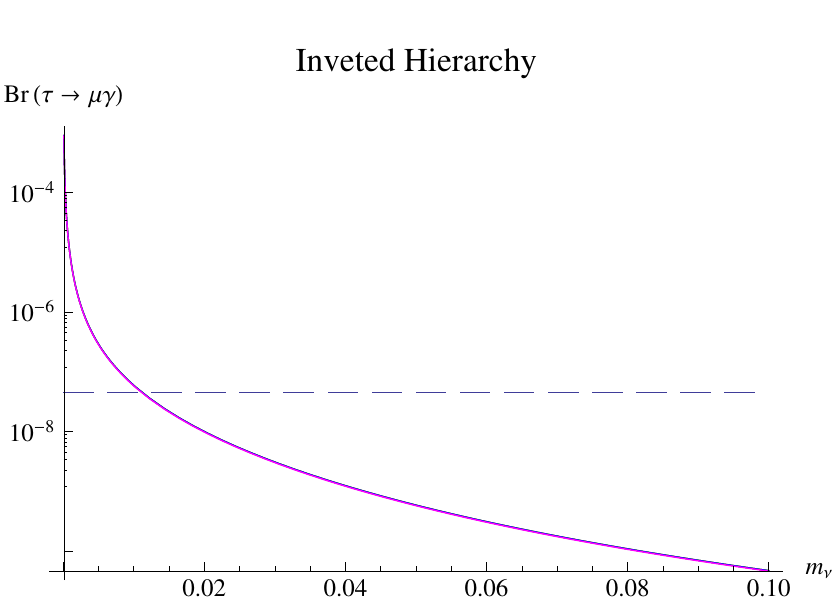}
\caption{
The lightest mass $m_{\nu}$ dependence of 
 branching ratios of $\mu\to e\gamma$, $\tau\to e\gamma$,
 and $\tau\to \mu\gamma$ with 
 $\alpha_{\nu}=1/4\pi$ and $m_{h^{+}}=1$ TeV. 
Red line shows 
 $\theta_{}=0$, 
 green line 
 $\theta_{}=0.001$, 
 blue line 
 $\theta_{}=0.01$, and 
 purple line 
 $\theta_{}=0.1$. 
Dashed line corresponds to each experimental bound. 
}
\label{fig2}
\end{center}
\end{figure}

Here, let us consider a possibility 
 that our model can generate enough large muon anomalous
 magnetic moment which is measured in 
 experiments\cite{Prades:2009qp} as 
\begin{equation}
\Delta a_{\mu} =(25.5\pm 8.0)\times 10^{-10}. 
\label{muona}
\end{equation}
Assuming $|m_{h^\pm}|\simeq |M_M|$, 
 the muon anomalous magnetic moment is given by\cite{Ma:2001mr} 
\begin{eqnarray}
\Delta a_{\mu} 
&\simeq &-\sum_{i}\frac{(y^{\nu}_{i \mu})^2}{12(4\pi)^{2}}
\frac{m^{2}_{\mu}}{m^{2}_{h^{+}}} =
 \frac{\alpha_{\nu}m^{2}_{\mu}}{96\pi
 m^{2}_{h^{+}}}\frac{(1+\sqrt{3}\sin\theta_{13})^{2}m_{1}+(\sqrt{3}-\sin\theta_{13})^{2}
m_{2}+4\cos^{2}\theta_{13}m_{3}}{m_{\nu}}\nonumber \\
&\simeq&
-\frac{\alpha_{\nu}m^{2}_{\mu}}{96\pi m^{2}_{h^{+}}}\frac{m_{1}+3m_{2}+4m_{3}}{m_{\nu}},
\end{eqnarray}
where we use $\theta_{12}=\pi/6,\theta_{23}=\pi/4$ 
 ($\theta_{13}=0$) 
 in the first (second) line.  
Unfortunately, the sign is opposite from Eq.(\ref{muona}),
 so that our model can not induce the deviation. 
This situation might be changed in the supersymmetric extension\cite{HH}.


\subsection{Majorana nature of neutrinos}

An idea of the model we suggested is similar 
 to the model\cite{{Ma}, {Nandi}, {Davidson:2009ha}}, 
 but 
 the biggest different point is that light active 
 neutrinos are Majorana particles
 in our model. 
Are there experimental predictions of Majorana natures for
 these active neutrinos?

One is a 
 neutrinoless double beta decay, 
 which never occur in case of Dirac neutrinos. 
The phenomenological analyses in this paper 
 used the neutrino oscillation data and 
 $\leq {\mathcal O}(0.1)$ eV absolute mass of neutrinos from 
 cosmology\cite{WMAP}. 
By using them, 
 prediction about neutrinoless double beta decay
 is obtained. 
Taking vanishing Majorana CP phases, for simplicity,
 it is given by 
 $\langle m_{\beta\beta}\rangle=
 \frac{3}{4}\cos^{2}\theta+\frac{1}{4}m_{2}\cos^{2}\theta+
 m_{3}\sin^{2}\theta$ by using  
 $\theta_{12}=\pi/6, \theta_{23}=\pi/4$. 
When we take the lightest mass as 0.1 eV 
 (0.01 eV, 0.001 eV), 
 NH shows 
 $\langle m_{\beta\beta}\rangle=$
 0.10 eV (0.011 eV, 0.0030 eV), and 
 IH 
 $\langle m_{\beta\beta}\rangle=$
 0.11 eV (0.049 eV, 0.048 eV), respectively.\footnote{ 
Even if we take into account of finite Majorana phases, 
 the magnitude does not increase. }
It is consistent with today's experimental 
 bound, $\langle m_{\beta\beta}\rangle<0.1$ eV\cite{PDG}. 

Anyhow, above results are obtained from the current 
 neutrino oscillation data, and they are not specific predictions from  
 our model, TeV-scale seesaw from multi-Higgs model. 
Are there any direct evidences in collider experiments of 
 our model? 
One of the important motivations for our model is detective new physics 
 at TeV-scale, and it is the 
 reason why we set TeV-scale for
 right-handed neutrinos. 

In a high energy collider experiments, 
 there is a chance of direct production of
 right-handed neutrinos.  
For example, in a linear collider, 
 there are T-channel processes of charged Higgs exchange 
 $e^+e^- \to 2N$, 
 $e^+e^- \to 2N \gamma$, 
 and so on. 
The first is missing event,
 and the latter 
 is a single photon event which can be detected clearly. 
The decay channels of $N$ are also interesting, since
 it can produce ($S$-originated) singlet scalars with lepton number. 
We will show detailed analyses in the next paper\cite{HH}.

\section{Summary and discussions}

We have proposed 
 a simple model for Majorana neutrino case, 
 which is a 
 renormalizable model with minimal
 extension of the SM which appears entirely below the TeV-scale.  
Tiny Majorana neutrino masses are obtained  
 through a TeV-scale type I seesaw mechanism 
 without requiring tiny Yukawa couplings. 
This model contains 
 extra one Higgs singlet, and 
 one Higgs doublet 
 with a tiny VEV.  
As for 
 extending a Higgs sector, 
 there are 
 constraints in general, which are consistency of 
 electroweak precision data
 and 
 absence of large FCNCs. 
In our model, 
 both two constraints are 
 satisfied 
 since the extra doublet only has a Yukawa interaction 
 with lepton doublets and right-handed neutrinos, and 
 their masses are heavy 
 enough to suppress FCNCs
 although its VEV is of order 0.1 eV. 
The extra Higgs doublet yields a neutral scalar and a
 neutral pseudo-scalar, and 
 a charged Higgs particles, 
 which can 
 provide collider signatures. 
This charged Higgs can contribute to the lepton flavor 
 violating processes. 
The extra singlet produces TeV-scale Majorana masses of right-handed 
 neutrinos, and yields a neutral scalar and a neutral pseudo-scalar 
 with a lepton number. 
Other phenomenology have also been represented such as 
 the charged Higgs decay processes depending on 
 the particle mass spectra. 
Notice that the decay of the charged Higgs to quarks and 
 charged leptons are strongly suppressed due to absence of 
 direct interactions among them, which 
 is one of different points from usual 
 two Higgs double models.  
Since active light neutrinos are Majorana particles, 
 this model predicts a neutrinoless double beta decay.

Finally, we give a comment. 
The supersymmetric extension can be also achieved 
 by introducing small magnitude of
 $A$-terms, which is expected to be induced 
 some supersymmetry breaking scenarios. 
However, for the suitable gauge coupling unification 
 we should introduce extra colored Higgs particles. 
In this case, we should introduce baryon number symmetry 
 to avoid rapid proton decay.


\vspace{1cm}

{\large \bf Acknowledgments}\\

\noindent
We thank M. Tanimoto, S. Matsumoto, S. Kanemura, 
 G. C. Cho, O. Seto, M. Tanaka, and K. Tsumura for useful and helpful discussions. 
This work is partially supported by Scientific Grant by Ministry of 
 Education and Science, Nos. 20540272, 20039006, and 20025004.

\appendix

\section{Conditions of bonded below potential}

We show 
 conditions of Higgs potential to be bonded below. 
To obtain the conditions,
 we do not need to take into account 
 mass terms and three-point interactions of Higgs fields. 
Thus, we must only take the following interactions,  
\begin{eqnarray} 
V &\sim &
\lambda_{1}h^{4}+\lambda_{2}h^{4}_{\nu}+(\lambda_{3}+\lambda_{4})
 h^{2}h^{2}_{\nu} 
+\lambda_{s}s^{4}+\lambda_{H}h^{2}s^{2}+\lambda_{H_{\nu}}h^{2}_{\nu}s^{2}. 
\end{eqnarray}
It is rewritten as 
\begin{eqnarray}
V&\sim & \frac{1}{2}(\lambda_{1}h^{4}+\lambda_{2}h^{4}_{\nu})+
 \frac{1}{2}(\lambda_{1}h^{4}+\lambda_{s}s^{4})+\frac{1}{2}
 (\lambda_{2}h^{4}_{\nu}+\lambda_{s}s^{4})
+(\lambda_{3}+\lambda_{4})h^{2}h^{2}_{\nu}+\lambda_{H}h^{2}s^{2}+
 \lambda_{H_{\nu}}h^{2}_{\nu}s^{2}, \nonumber \\
&>&
(\sqrt{\lambda_{1}\lambda_{2}}+\lambda_{3}+\lambda_{4})h^{2}h^{2}_{\nu}
+(\sqrt{\lambda_{1}\lambda_{s}}+\lambda_{H})h^{2}s^{2}
+(\sqrt{\lambda_{2}\lambda_{s}}+\lambda_{H_{\nu}})h^{2}_{\nu}s^{2},
\label{a71}
\end{eqnarray}
where we use bonded below for each field's direction, 
\begin{eqnarray}
\lambda_{1}>0, \;\;\;
\lambda_{2}>0, \;\;\;
\lambda_{s}>0. 
\label{a72}
\end{eqnarray}
Then, a necessary and sufficient condition 
 of bounded below potential is 
 that all coefficients in (\ref{a71}) are real and positive, 
 which are given by 
\begin{eqnarray}
\sqrt{\lambda_{1}\lambda_{2}}>-\lambda_{3}-\lambda_{4}, \;\;\;  
\sqrt{\lambda_{1}\lambda_{s}}>\lambda_{H}, \;\;\; 
\sqrt{\lambda_{2}\lambda_{s}}>\lambda_{H_{\nu}}. 
\label{a73}
\end{eqnarray}
Therefore,
 the condition 
 of bounded below is given by 
 Eqs.(\ref{a72}) and (\ref{a73}).

\section{Higgs interactions}

Here, we summarize Higgs interactions below the energy scale of 
 Higgs VEVs with an assumption of CP invariance 
 in the Higgs sector. 
 We denote scalars, pseudo-scalars, and charged Higgs 
 as 
\begin{eqnarray}
R_{i}=
\left(
\begin{array}{c}
H_{S}\\
h_{0}\\
H_{0}
\end{array}
\right), \;\;\;
P_{i}=
\left(
\begin{array}{ccc}
A_{S}\\
G^{0}\\
A_{0}
\end{array}
\right) \;\;\;
\end{eqnarray}

\noindent
{\bf $\cdot$ 3-points interactions of $R_{i}h^{+}h^{-}$}:

\noindent
Interactions are given by 
\begin{eqnarray}
&((2\lambda_{H_{\nu}}s\cos^{2}\beta_{3}-\mu\sin2\beta_{3}-2\lambda_{H}s\sin^{2}\beta_{3})V_{1i}
+(2\lambda_{3}h\cos^{2}+\lambda_{4}h_{\nu}\sin2\beta_{3}+2\lambda_{1}h\sin^{2}\beta_{3})V_{2i}
 \nonumber \\
&+(2\lambda_{2}h_{\nu}\cos^{2}\beta_{3}+\lambda_{4}h\sin2\beta_{3}+2\lambda_{3}h_{\nu}\sin^{2}\beta_{3})V_{3i})
R_{i}h^{+}h^{-}
\end{eqnarray}
where
 $V_{ij}$ is a mixing matrix defined in Eqs.(\ref{223}) and (\ref{224}).

\noindent
{\bf $\cdot$ 4-points interactions of $R_{i}R_{j}h^{+}h^{-}$ and
 $P_{i}P_{j}h^{+}h^{-}$}: 

\noindent
They are given by 
\begin{equation}
\sum_{m,n}(V^{\dagger})_{im}O_{mn}V_{nj}R_{i}R_{j}h^{+}h^{-}
+\sum_{m,n}(V^{\prime\dagger})_{im}O_{mn}V^{\prime}_{nj}
 P_{i}P_{j}h^{+}h^{-},
\end{equation}
where
\begin{eqnarray}
&O_{mn}=
\left(
\begin{array}{ccc}
\lambda_{H_{\nu}}\cos^{2}\beta_{\nu}-\lambda_{H}\sin^{2}\beta_{3} &0 &0
\\
0&\lambda_{3}\cos^{2}\beta_{3}+\lambda_{1}\sin^{2}\beta_{3}&\frac{\lambda_{4}}{2}\sin2\beta_{3}
\\
0&\frac{\lambda_{4}}{2}\sin2\beta_{3}&\lambda_{2}\cos^{2}\beta_{3}+\lambda_{3}\sin^{2}\beta_{3}
\end{array}
\hspace{-2mm} \right)
\end{eqnarray}
and $V^{\prime}$ is a mixing matrix for imaginary part 
 defined in Eqs.(\ref{226}) and (\ref{227}). 

\noindent
{\bf $\cdot$ 4-points interactions of $h^{+}h^{-}h^{+}h^{-}$}:

\noindent
They are give by
\begin{equation}
\bigl(\frac{1}{2}\lambda_{1}\sin^{4}\beta_{3}+\frac{1}{2}\lambda_{2}\cos^{4}\beta_{3}+(\lambda_{3}+\lambda_{4})\cos^{2}\beta_{3}\sin^{2}\beta_{3}\bigr)
h^{+}h^{-}h^{+}h^{-}. 
\end{equation}

\noindent
{\bf $\cdot$ 3-points interactions of neutral Higgs}: 

\noindent
They are given by 
\begin{eqnarray}
\sum_{l,m,n}T_{lmn}V_{li}(V^{\dagger})_{jm}V_{nk}R_{i}R_{j}R_{k}
+\sum_{l,m,n}T^{\prime}_{lmn}V_{li}(V^{\prime\dagger})_{jm}V^{\prime}_{nk}R_{i}P_{j}P_{k}, 
\end{eqnarray}
where $T_{lmn}$ is a  symmetric tensor 
 as follows
\begin{eqnarray}
&T_{111}=-2\lambda+4\lambda_{S}s, \;\;
T_{112}=-\frac{2}{3}\lambda_{H}h, \;\;
T_{113}=\frac{2}{3}\lambda_{H_{\nu}}h_{\nu},\;\;
T_{122}=-\frac{2}{3}\lambda_{H}s,\;\;
T_{123}=-\frac{\mu}{3}, \nonumber\\
&T_{133}=\frac{2}{3}\lambda_{S}s,\;\;
T_{222}=2\lambda_{1}h,\;\;
T_{223}=\frac{2}{3}(\lambda_{3}+\lambda_{4})h_{\nu},\;\;
T_{233}=\frac{2}{3}(\lambda_{3}+\lambda_{4})h,\;\;
T_{333}=2\lambda_{2}h_{\nu}, \nonumber 
\end{eqnarray}
with 
\begin{eqnarray}
T^{\prime}_{1jk}=
\left(
\begin{array}{ccc}
6\lambda+4\lambda_{S}s&0&0
\\
0&-2\lambda_{H}s&-\mu
\\
0&-\mu&2\lambda_{H_{\nu}}s
\end{array}
\right),\nonumber 
\\
T^{\prime}_{2jk}=
\left(
\begin{array}{ccc}
-2\lambda_{H}h&0&\mu
\\
0&2\lambda_{1}h&0
\\
\mu&0&2(\lambda_{3}+\lambda_{4})h
\end{array}
\right),\nonumber 
\\
T^{\prime}_{3jk}=
\left(
\begin{array}{ccc}
2\lambda_{H_{\nu}}h_{\nu}&-\mu&0
\\
-\mu&2(\lambda_{3}+\lambda_{4})h_{\nu}&0
\\
0&0&2\lambda_{2}h_{\nu}
\end{array}
\right). \nonumber 
\end{eqnarray}

\noindent
{\bf $\cdot$ 4-points interactions of neutral Higgs}: 

\noindent
They are given by 
\begin{eqnarray}
&&\sum_{m,n,s,t}X_{mnst}(V^{\dagger}_{mi})(V^{\dagger}_{jn})
V_{sk}V_{tl}R_{i}R_{j}R_{k}R_{l}
+\sum_{m,n,s,t}X^{\prime}_{mnst}(V^{\dagger}_{mi})
(V^{\prime\dagger}_{jn})V_{sk}V^{\prime}_{tl}R_{i}P_{j}R_{k}P_{l}
\nonumber \\
&&
+\sum_{m,n,s,t}X_{mnst}(V^{\prime\dagger}_{mi})
(V^{\prime\dagger}_{jn})V^{\prime}_{sk}V^{\prime}_{tl}P_{i}P_{j}P_{k}P_{l}, 
\end{eqnarray}
where 
$X_{1111}=\lambda_{S}$, 
$X_{\sigma(1122)}=-\frac{\lambda_{H}}{6}$, 
$X_{\sigma(1133)}=\frac{\lambda_{H_{\nu}}}{6}$, 
$X_{2222}=\frac{1}{2}\lambda_{1}$,
$X_{\sigma(2233)}=\frac{1}{6}(\lambda_{3}+\lambda_{4})$, 
$X_{3333}=\frac{1}{2}\lambda_{2}$,
\begin{eqnarray}
&
X^{\prime}_{11st}=
\left(
\begin{array}{ccc}
2\lambda_{S}&0&0\\
0&-\lambda_{H}&0\\
0&0&\lambda_{H_{\nu}}
\end{array}
\right) 
,\;\;
X^{\prime}_{22st}=
\left(
\begin{array}{ccc}
-\lambda_{H}&0&0\\
0&\lambda_{1}&0\\
0&0&\lambda_{3}+\lambda_{4}
\end{array}
\right)
\nonumber
\\
&
X^{\prime}_{33st}=
\left(
\begin{array}{ccc}
\lambda_{H_{\nu}}&0&0\\
0&\lambda_{3}+\lambda_{4}&0\\
0&0&\lambda_{2}
\end{array}
\right)
, \;\;
{\rm others}=0. \nonumber  
\end{eqnarray}


\end{document}